\documentclass[jkps,preprint,fleqn,showpacs,showkeys]{revtex4}
\usepackage[pdftex]{graphicx}
\usepackage{amssymb}
\usepackage{amsmath}
\usepackage{bm}
\begin{document}
\title[]{Effects of Structural Distortion Induced by Sc Substitution in LuFe$_2$O$_4$}
\author{Jinwon \surname{Jeong}}
\author{Han-Jin \surname{Noh}}
\email{ffnhj@jnu.ac.kr}
\thanks{Fax: +82-62-530-3369}
\affiliation{Department of Physics, Chonnam National University, Gwangju 500-757}
\author{Sung Baek \surname{Kim}}
\affiliation{Advancement for College Education Center, Konyang University, Chungnam, 320-711}

\date{\today}

\begin{abstract}
We have studied the correlation between the structural distortion and the electronic/magnetic properties in single-crystalline (Lu,Sc)Fe$_2$O$_4$ (Sc=0.05 and 0.3) by using X-ray diffraction (XRD), magnetic susceptibility, and X-ray absorption spectroscopy (XAS)/X-ray magnetic circular dichroism (XMCD) measurements.
The Rietveld structure analysis of the XRD patterns revealed that the Sc substitution induced an elongation of the FeO$_5$ bipyramidal cages in LuFe$_2$O$_4$ and increased the Fe$_2$O$_4$ bilayer thickness.
A non-negligible decrease in the ferrimagnetic transition temperature T$_C$ is observed in the magnetic susceptibility curve of the Sc=0.3 sample, but the XAS/XMCD spectra do not show any difference except for a small reduction of dichroism signals at the Fe$^{3+}$ absorption edge.
We interpret this suppression of T$_C$ to be the result of a decreased spin-orbit coupling effect in the Fe$^{2+}$ $e_{1g}$ doublet under $D_{3h}$ symmetry, which is induced by the weakened structural asymmetry of the FeO$_5$ bipyramids.

\end{abstract}




\maketitle

\section{Introduction}

The revived interest in multiferroic systems over the last decade has found various kinds of couplings between multiple long-range orders \cite{Cheong}.
One prototypical example for the variety is the ferroelectricity induced by Fe valence ordering in bilayered LuFe$_2$O$_4$ \cite{Ikeda}.
This system has mixed valent Fe cations with a one-to-one ratio of divalent to trivalent ions, so the average value of the Fe valences is Fe$^{2.5+}$.
At temperatures below $\sim$320 K, the trivalent ions and the divalent ions are ordered to a specific pattern that breaks inversion symmetry, thus inducing ferroelectricity.
This ferroelectricity induced by valence ordering is quite contrastive and unique in that the traditional origin of spontaneous electric polarization in solids is the so-called $d^0$-ness where transition-metal cations with no $d$-electrons are located at an off-centered position \cite{Khomskii}.
However, successive studies on this polar state have challenged the originally proposed charge-ordered state, and this issue is still under debate \cite{Xu, Ren, deGroot1, Niermann, Kambe, JLee}.
Even though whether LuFe$_2$O$_4$ has a polar state in a macroscopic range has not been definitely determined, the charge-ordered state within an Fe$_2$O$_4$ bi-layer is clearly polarized, so a concrete interaction study on the charge and spin orders is still important for understanding this system.

A magnetic transition occurs at T$_C\sim$240 K in this system, so multiple long-range orders, i.e., ferroelectric order and ferrimagnetic order, co-exist at temperatures below T$_C$.
The exact spin ordering pattern in the Fe$_2$O$_4$ hexagonal bilayer has been proposed based on the X-ray absorption spectroscopy (XAS) and X-ray magnetic circular dichroism (XMCD) studies \cite{Ko}.
The analysis of the XAS/XMCD spectra suggests that the Fe$^{2+}$ spins are ferromagnetically ordered by the spin-orbit-coupling while the Fe$^{3+}$ spins are antiferromagnetically ordered, possibly by the superexchange interaction.
The issue on the inter-bilayer magnetic coupling in LuFe$_2$O$_4$ is actually coupled to the question of whether this system is intrinsically ferroelectric or not \cite{deGroot2}.
Our previous neutron diffraction study on this issue suggested that the inter-bilayer coupling in LuFe$_2$O$_4$ was a mixture of ferromagnetic and antiferromagnetic ones, but that the ferromagnetic type was preferred \cite{Noh}.

The effects of A-site substitution on the Fe spin ordering pattern and on the magnetic transition temperature have been studied in Lu$_{1-x}$L$_x$Fe$_2$O$_4$ (L=Y or Er; $x$=0.1, 0.5)\cite{Noh} or in LFe$_2$O$_4$ (L=Er or Tm) \cite{DHKim}.
Partial substitution induces a flattening of the hexagonal unit cell and a frustration of Fe$^{3+}$ spin ordering, but the magnetic transition temperature and the $\sqrt{3}\times\sqrt{3}$ superstructure formation are not affected very much.
A full substitution study also shows behaviors very similar to those of LuFe$_2$O$_4$, except for indications of extra magnetic phases at a temperature lower than T$_C$ $\sim$240 K.
This result is a little bit disappointing if we want to realize a room-temperature multiferroic system other than BiFeO$_3$ \cite{SLee}.
However, the previous study covers only a partial range of the lattice distortion, i.e., a flattening-type distortion of the unit cell, so a complementary study on the effect of lattice elongation is required to fully understand the exact Fe spin interactions in this class of materials.
An experimental investigation toward this complementary direction can be performed by using an appropriate choice of substituted cations with ionic radii smaller than that of the Lu$^{3+}$ ions.

In this research, we focused on the lattice elongation effect in LuFe$_2$O$_4$ and scrutinized the correlation between the electronic/magnetic properties and the lattice distortion by performing X-ray diffraction (XRD), magnetic susceptibility, XAS, and XMCD measurements on Sc-substituted LuFe$_2$O$_4$ single-crystalline samples grown by using a floating zone method.
The XRD pattern analysis shows that Sc substitution induces an elongation of the hexagonal unit cell in LuFe$_2$O$_4$, together with an elongation of the FeO$_5$ bipyramidal cages in the Fe$_2$O$_4$ bilayers.
For the Sc=0.3 sample, a discernible decrease ($\sim$30 K) in the magnetic transition temperature is observed in the magnetic susceptibility curve, but no considerable changes in the XAS/XMCD spectra are observed.
The decrease in T$_C$ implies the existence of a correlation between the structural distortion of the Fe$_2$O$_4$ bilayers and the Fe$^{2+}$-Fe$^{2+}$ magnetic interaction strength.
We interpret this to be a result of the decreased spin-orbit coupling effect in the Fe$^{2+}$ 3$d$ $e_{1g}$ doublet, which is induced by the weakened structural asymmetry of the FeO$_5$ bipyramids.

\section{Experiments}
High-quality polycrystalline Lu$_{1-x}$Sc$_x$Fe$_2$O$_4$ ($x$=0.05 and 0.3) was synthesized by using the solid state reaction method.
Stoichiometric amounts of high-purity ($\geq$ 99.99\%) dehydrated Lu$_2$O$_3$, Sc$_2$O$_3$, and Fe$_3$O$_4$ powders were weighed and mixed with a pestle and mortar.
The mixtures were fired two times for feed rods under a flow of a CO/CO$_2$ gas mixture at 1200$^\circ C$ in order to achieve the right oxygen stoichiometry \cite{Iida}.
The single crystals were grown with the feed rods by using the floating-zone method with an optical furnace in the laboratory of Pohang Emergent Materials.
During the crystal growth process, the flow of the CO/CO$_2$ gas mixture was maintained at atmospheric pressure.
The single phase of the samples in this work was checked by using XRD at each step.

Magnetization measurements were performed using a commercial magnetometer with a superconducting quantum interference device.
The X-ray absorption and X-ray magnetic circular dichroism experiments were performed at the 2A elliptically-polarized undulator beamline in the Pohang Light Source.
The samples were cleaved {\it in situ} by using the so-called top-post method at a pressure of $\sim$3$\times$10$^{-10}$ Torr.
The polarization-dependent X-ray absorption spectra were obtained at 210 and 300 K, respectively, with $\sim$98\% linearly-polarized light at 0$^\circ$ and 60$^\circ$ beam incidence to the surface normal.
For the XMCD measurements, a 0.6 T magnetic field was applied to the samples to align the magnetic moments at 210 K, and the incident photon angle was 22.5$^\circ$ relative to the magnetic field's direction.
The absorption spectra with an energy resolution of $\sim$0.15 eV were acquired in a total electron yield mode and were normalized to the incident photon flux.

\section{Results and Discussion}

\begin{figure}
\includegraphics[width=12.0cm]{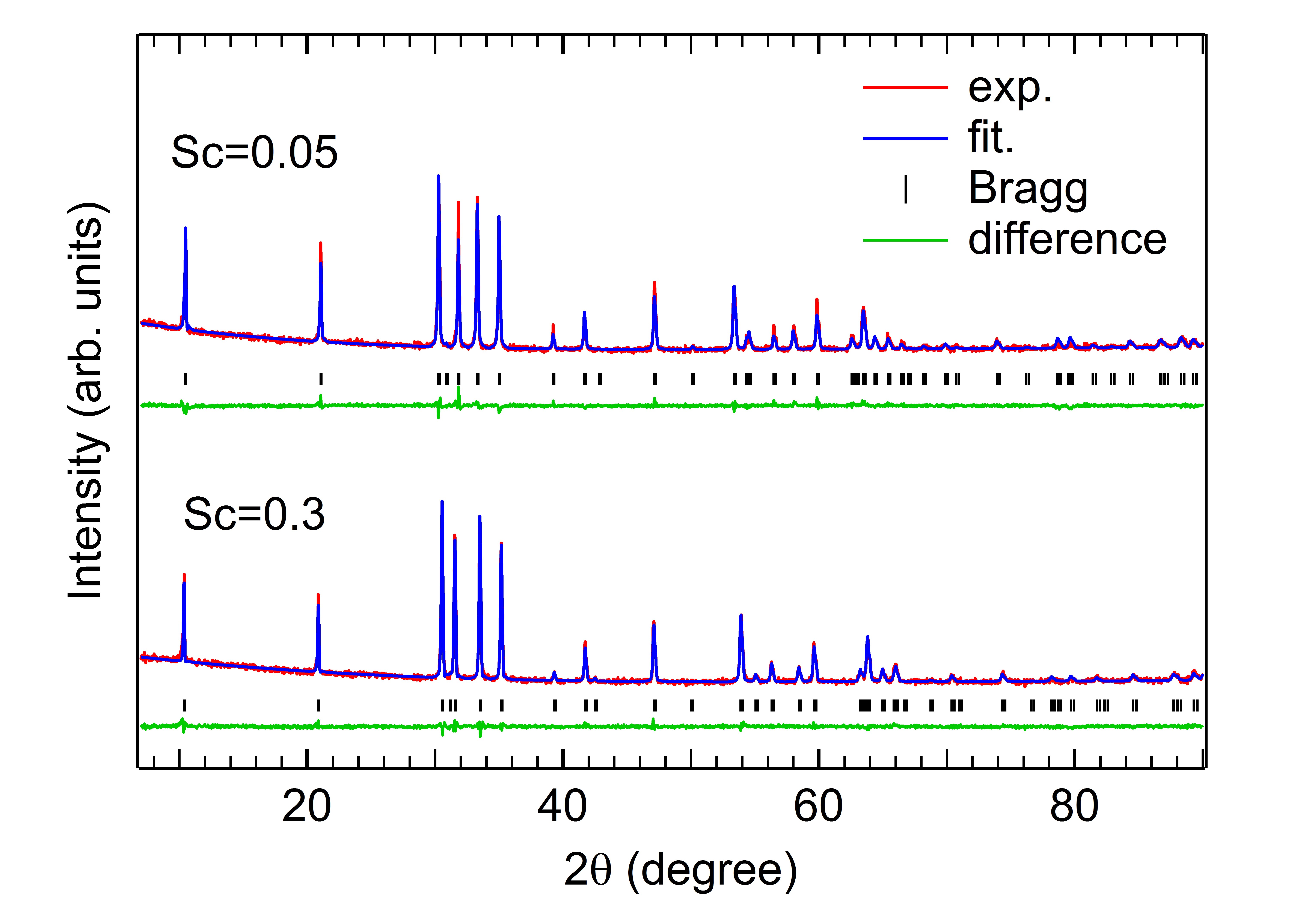}
\caption{(Color online) X-ray diffraction patterns of hexagonal bilayered Lu$_{1-x}$Sc$_x$Fe$_2$O$_4$ ($x$=0.05 and 0.3) at 300 K.
}\label{XRD}
\end{figure}

\begin{figure}
\includegraphics[width=10.0cm]{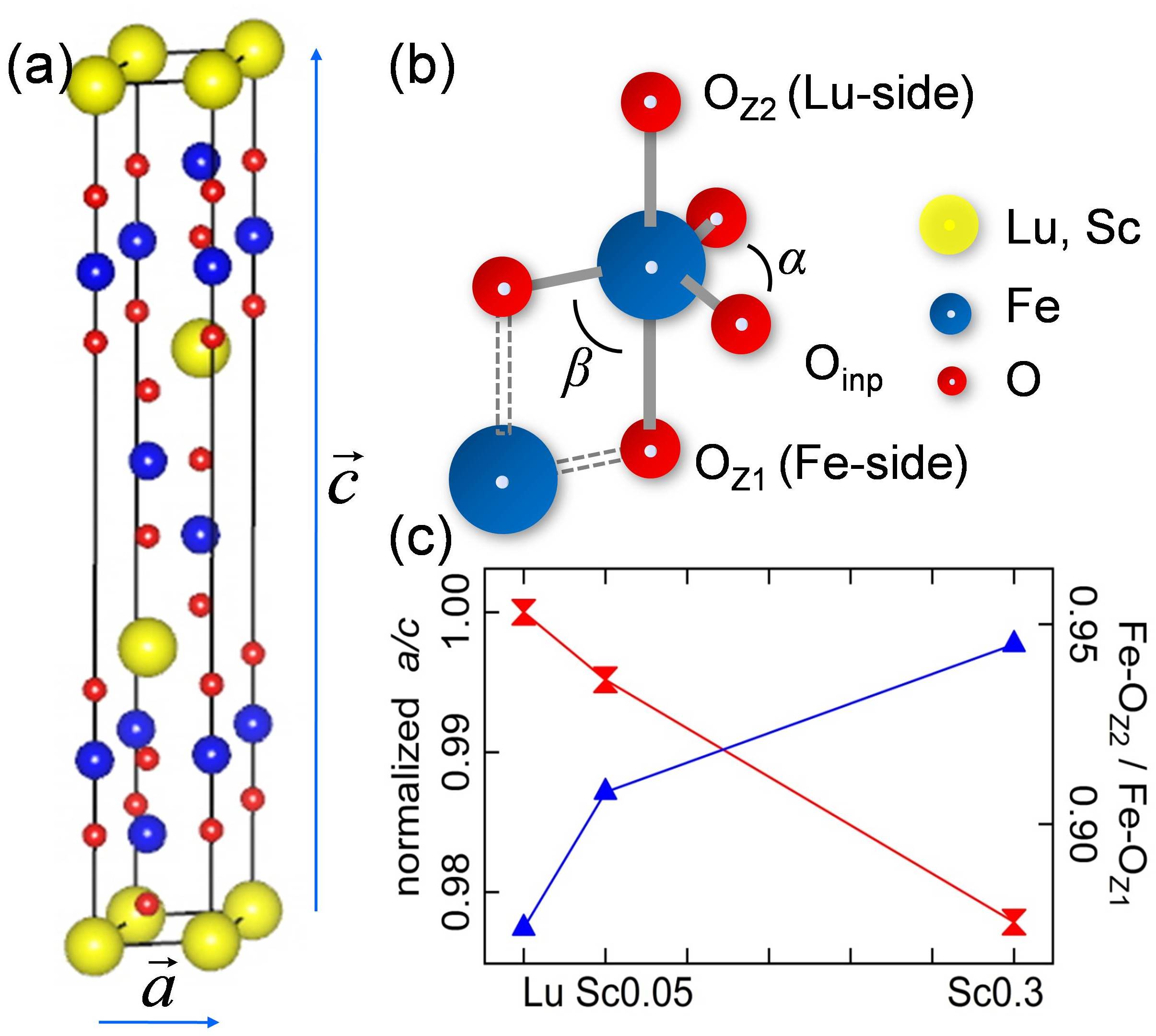}
\caption{(Color online) (a) Crystal structure of bilayered LuFe$_2$O$_4$.
(b) Bipyramidal FeO$_5$ cage.
(c) Normalized $a/c$ ratio vs. Sc concentraion (red sandwatchs, left ornidate), and  the bond length ratio of Fe-O$_{z2}$ to Fe-O$_{z1}$ (blue triangles, right ordinate).
}\label{crystal}
\end{figure}

Figure 1 shows the XRD patterns of our Sc-substituted LuFe$_{2}$O$_4$ with a Rietveld analysis result.
For each XRD pattern, the blue curve is the fitting profile, the black bars are the positions of the Bragg peaks, and the green line is the differences between the observations and the fittings.
The two patterns look very similar except for small shifts in the positions of the peaks, implying that Sc substitution induces changes in the lattice parameters without inducing changes in the symmetry of the crystal structure.
Because the crystal structure of the mother compound LuFe$_2$O$_4$ is rhombohedral ($R\overline{3}m$) at temperatures above the charge-ordering temperature (T$_{CO}$=320 K), we refined the crystal structure of Sc-substituted LuFe$_2$O$_4$ in the same symmetry by using Fullprof \cite{fullprof}, mainly focusing on the distortion of the FeO$_5$ bipyramidal cages as a function of the Sc concentration.
The unit cell of the crystal structure is shown in Fig. 2(a).
As the Sc concentration increases, the lattice constant $a$ decreases, but $c$ increases with a small amount of decrease in unit cell volume.
In the local structure, the average bond length of Fe-O$_{inp}$ gets shorter, but that of Fe-O$_{apical}$ gets longer, so the FeO$_5$ bipyramids are elongated.
One noticeable thing is that the asymmetry of the bipyramids gets weaker with increasing Sc substitution.
The FeO$_5$ cages have two kinds of apical oxygen, O$_{z1}$ (Fe side) and O$_{z2}$ (Lu side), as shown in Fig. 2(b).
The length of Fe-O$_{z1}$ is $\sim$10\% longer than that of Fe-O$_{z2}$ in the Sc=0.05 sample, but Fe-O$_{z1}$ is only $\sim$5\% longer than that of Fe-O$_{z2}$, together with the $\sim$9\% increase in the absolute bond length in the Sc=0.3 sample.
The unit cell elongation remains approximately linear up to Sc=0.3 as shown in Fig. 2(c).
In comparison with the bond length change, the major bond angles of O-Fe-O, $\alpha$ and $\beta$, change only within an $\sim$1\% range.
Detailed information on the lattice elongation is available in Table~\ref{table1}.

\begin{table}
\caption{Lattice constants and cell parameters for Lu$_{1-x}$Sc$_x$Fe$_2$O$_4$ ($x$=0.05 and 0.3).}
\begin{ruledtabular}
\begin{tabular}{cccccccccc}
Sc & $a$ (=$b$) & $c$ & V & Fe-O$_{inp}$ & Fe-O$_{z1}$ & Fe-O$_{z2}$ & $\alpha$ & 180$^\circ -\beta$ & $\chi^2$ \\
($x$)& [\AA] & [\AA] & [\AA$^3$] & [\AA]  & [\AA]  & [\AA]  & [deg] & [deg] & \\
\colrule
 0.05 & 3.4318 & 25.306 & 258.10 & 1.996(2) & 2.126(18) & 1.930(3) & 118.51 & 97.1 & 2.04 \\
 0.3  & 3.3992 & 25.508 & 255.25 & 1.986(17) & 2.233(11) & 2.110(2) & 117.66 & 98.9 & 1.67 \\
\end{tabular}
\end{ruledtabular}
\label{table1}
\end{table}

\begin{figure}
\includegraphics[width=6.0cm]{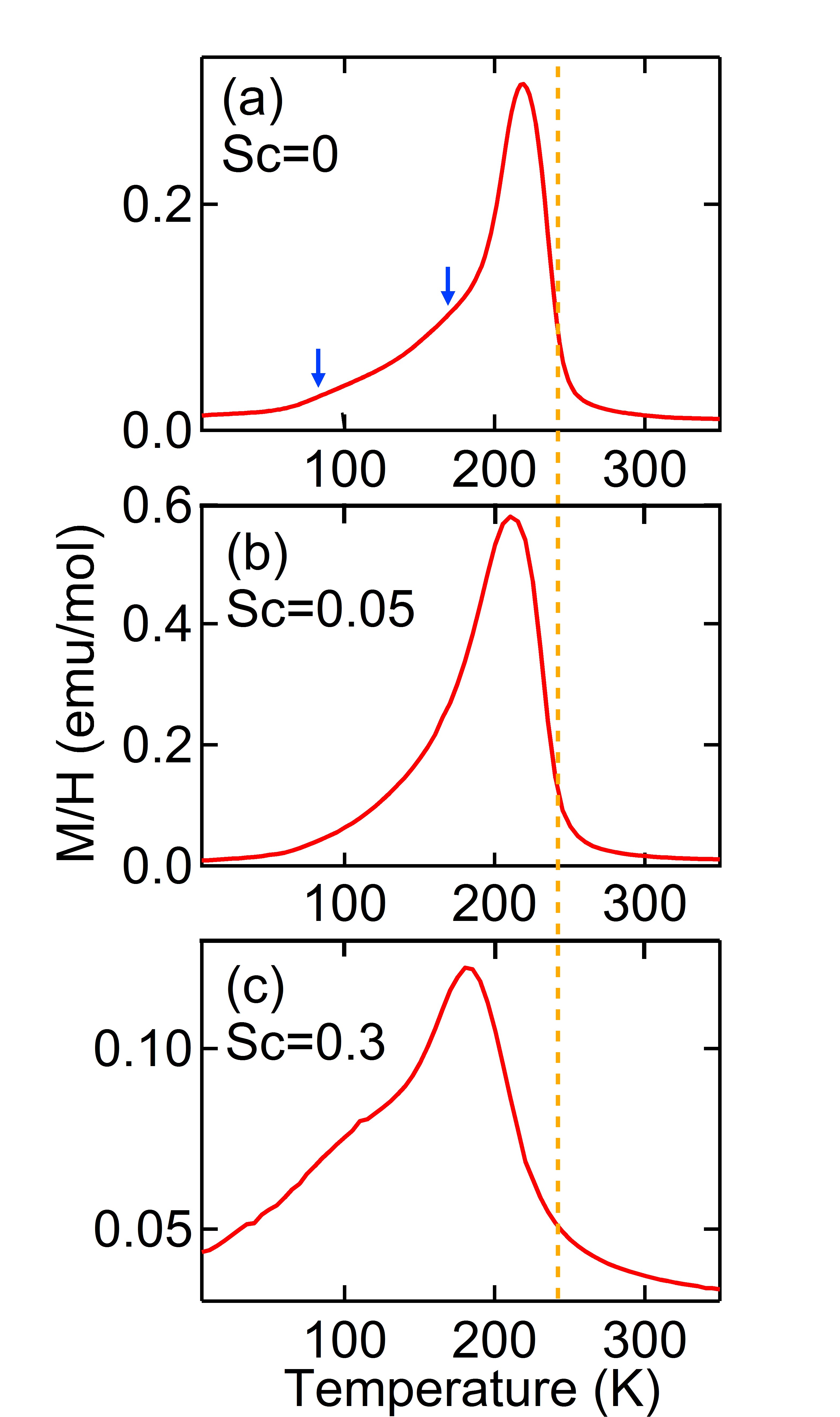}
\caption{(Color online) Temperature dependence of the magnetic susceptibility of Lu$_{1-x}$Sc$_x$Fe$_2$O$_4$ ($x$=0.05 and 0.3).
}\label{MT}
\end{figure}

Figure 3 shows the magnetic susceptibility curves measured in the zero-field-cooled mode under an external field of H=0.2 T as a function of sample temperature for Sc-substituted LuFe$_2$O$_4$.
From top to bottom, the Sc concentration increases from Sc=0.0 to Sc=0.3, but the susceptibility curves do not show considerable differences in shape.
They are quite similar to that of pristine stoichiometric LuFe$_2$O$_4$ reported so far.
The magnetization starts to develop at $\sim$240 K, reaches its maximum value at $\sim$210 K, then decreases down to 4 K.
The magnetic transition at T$_C$$\sim$240 K is known to be a ferrimagnetic one \cite{Ko} and to be common in all hexagonal RFe$_2$O$_4$ (R=rare-earth metal elements) systems.
A clear understanding of the decreasing magnetization at temperatures below ~210 K has not been established yet.
The magnitude and the shape of the curve in this temperature range severely depend on the sample preparation conditions and the measuring mode, so several possible scenarios such as the local antiferromagnetic stacking model \cite{Christianson},  the collective freezing model \cite{Wu}, and the charge fluctuation model \cite{XuPRL}, have been proposed, but no clear converging view exists yet.
Also, multiple magnetic transitions are reported at $\sim$175, $\sim$135, and $\sim$80 K in an oxygen stoichiometric study and a thermal treatment study \cite{Christianson, Wu, Patankar}, two of which are barely seen in Fig. 3(a), as indicated by the arrows.
Because these low-temperature transitions are reported to become prominent for non-stoichiometric samples \cite{Wang}, the shape of our susceptibility curves indicates that the Sc-substituted samples are close to stoichiometric.
All our samples but Sc=0.3 show magnetic transitions at T$_C$$\sim$240 K.
The Sc=0.3 sample shows a considerable lowering of the magnetic transition temperature (T$_C$$\sim$210 K).
Although the transition temperature is not increased, but decreased, by Sc substitution in LuFe$_2$O$_4$, this is a quite noticeable result because the transition temperature in this class of materials is known to be quite robust.
Because Sc substitution induces an elongation of the FeO$_5$ bipyramids, this implies that the Fe-O$_{z1/z2}$ length could be the dominant parameter for determining T$_C$.

\begin{figure}
\includegraphics[width=12.0cm]{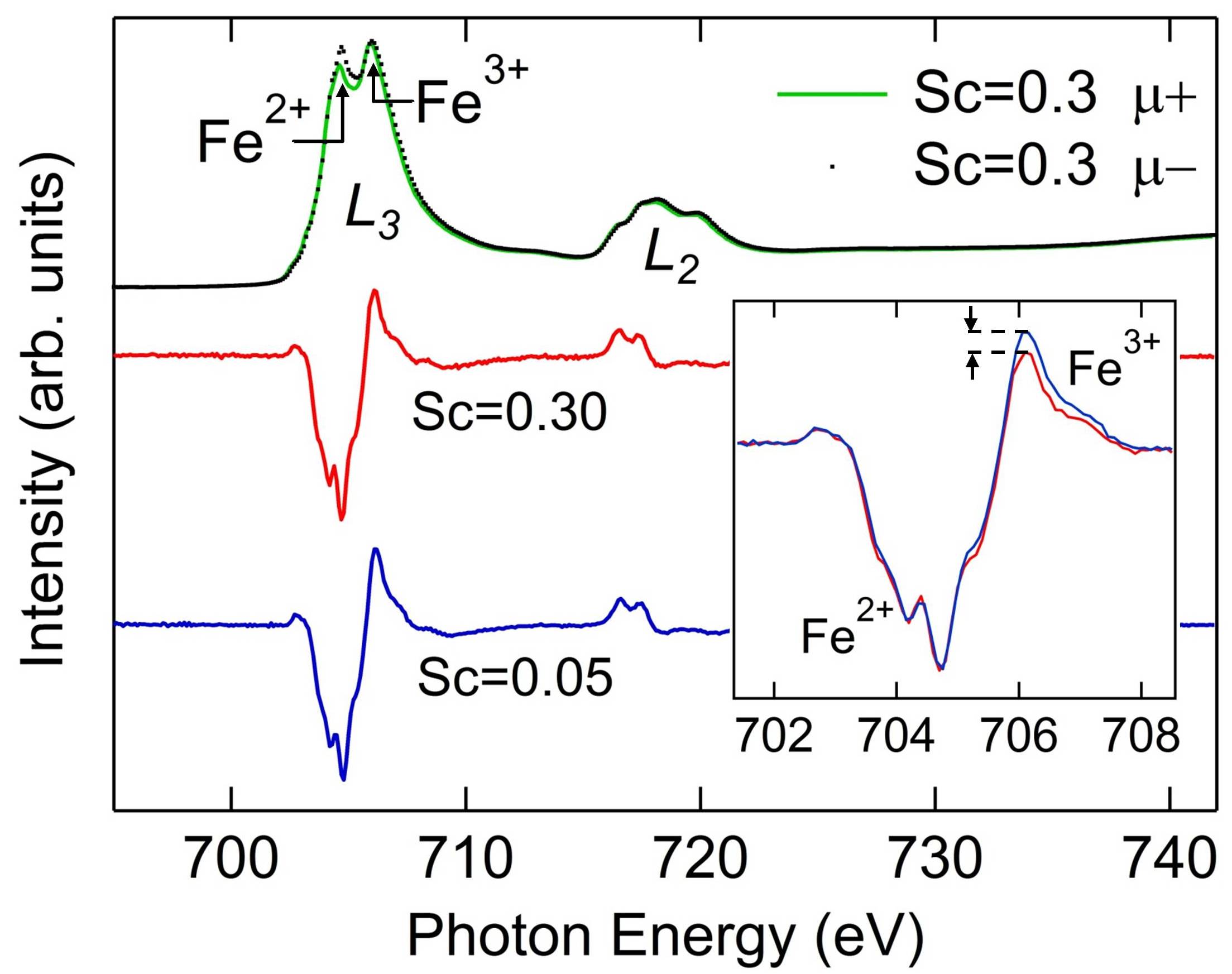}
\caption{(Color online) Fe 2$p$ XAS spectra (green/black) of Lu$_{0.7}$Sc$_{0.3}$Fe$_2$O$_4$ at 210 K, and Fe 2$p$ XMCD spectra of Lu$_{1-x}$Sc$_x$Fe$_2$O$_4$ ($x$=0.05 (blue) and 0.3 (red)).
}\label{XMCD}
\end{figure}

The effects of the Sc substitution on the Fe valence are checked by using Fe 2$p$ XAS/XMCD.
In Fig.~4 top, the absorption spectra with the photon helicity vector $\vec{k}$ parallel (solid line) and antiparallel (dotted line) to the sample magnetization $\vec{M}$ at the Fe $L$-edge are displayed.
Each spectrum shows the spin-orbit split $L_3$ and $L_2$ structure at $\sim$707 and $\sim$719 eV, respectively.
In the $L_3$ structure, two peaks with similar intensities are clearly resolved, as designated by Fe$^{2+}$ and Fe$^{3+}$, which indicates Fe$^{2+}$ and Fe$^{3+}$ ions co-exist in a 1:1 ratio in the Sc=0.3 sample.
The XAS spectrum for the Sc=0.05 sample is not displayed here, but it is statistically identical to that of Sc=0.3 sample.
Not only the XAS spectra but also their dichroism signals look quite similar to each other, as can be checked in Fig.~4.
From bottom to top, the XMCD spectrum for Sc=0.05 (blue) and 0.3 (red) are displayed, respectively.
The large downward peak at the Fe$^{2+}$ white line means that the net spin moment of Fe$^{2+}$ (S=2) ions is parallel to the applied magnetic field while the small upward peak at the Fe$^{3+}$ white line means that the net spin moment of Fe$^{3+}$ (S=5/2) is antiparallel to the applied magnetic field.
Thus, the XMCD signal indicates that the Fe spins are ferrimagnetically aligned.
Also, the dichroism weight of S=5/2 being smaller than that of S=2 gives a constraint that not all S=5/2 spins are aligned parallel to one another.
The dichroism peak weight of S=5/2 looks to be almost unaffected by the Sc concentration.
In the inset of Fig.~4, we overlap the XMCD spectra to compare the dichroism signals, and only a small reduction of the dichroism peak weight of Fe$^{3+}$ ions for Sc=0.3 is discerned.
Thus, the effects of Sc substitution on the Fe$^{3+}$ spin alignment can be regarded as negligible.
The concrete Fe spin alignment extracted from the XMCD spectra depends on the charge-ordering model proposed for this system.
If based on the $\sqrt{3}\times\sqrt{3}$ superstructure originally suggested by Ikeda \emph{et al} \cite{Ikeda}, three Fe$^{2+}$ spins and one Fe$^{3+}$ spin are parallel to the sample magnetization vector $\vec{M}$ and two Fe$^{3+}$ spins are antiparallel to $\overrightarrow{M}$ \cite{Ko, Noh}.
Our XMCD spectra indicate that the magnetic ordering pattern is not affected even though the ferrimagnetic transition temperature is lowered by $\sim$30 K in the Sc=0.3 sample.

\begin{figure}
\includegraphics[width=10.0cm]{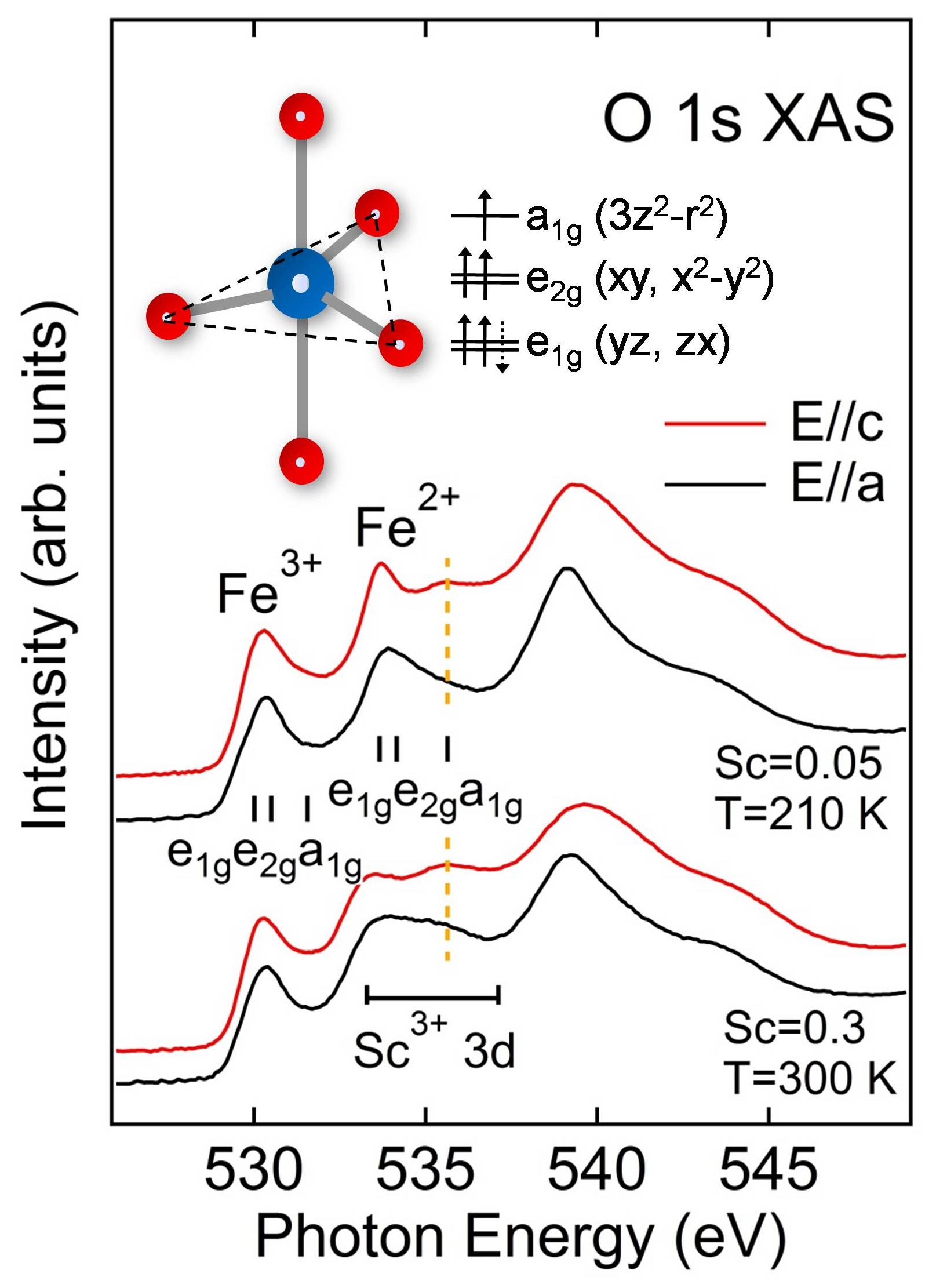}
\caption{(Color online) Polarization-dependent O $K$ XAS spectra of Lu$_{1-x}$Sc$_x$Fe$_2$O$_4$ ($x$=0.05 and 0.3). (Inset) Schematic diagram of the Fe 3$d$ level splittings under a bipyramidal crystal electric field.
}\label{OXAS}
\end{figure}

Figure~5 shows the polarization-dependent O K-edge XAS spectrum of each Sc-substituted sample.
The spectra commonly show three prominent features in the range of  530 $\sim$ 550 eV, which correspond to Fe$^{3+}$ 3$d$, Fe$^{2+}$/Sc$^{3+}$ 3$d$, and Lu$^{3+}$ 5$d$, 6$sp$/Fe/Sc 4$sp$ unoccupied states, respectively.
As in the case of XAS/XMCD at the Fe $L$-edge, they do not show any prominent dependence on the Sc concentration except for the overlying Sc 3$d$ states in the region of the Fe$^{2+}$ 3$d$ states ($\sim$536 eV).
Under the crystal electric field produced by a bi-pyramidal FeO$_5$ cage, the Fe 3$d$ states split into two doublets, $e_{1g}$ ($zx$, $yz$),  $e_{2g}$ ($xy$, $x^2-y^2$), and one singlet, $a_{1g}$ ($3z^2-r^2$), as shown in the schematic of Fig.~5 \cite{Cho}.
If we consider the dipole selection rule that dominates the absorption process, the O $K$-edge absorption coefficient for the Fe 3$d$ unoccupied states is proportional to the O 2$p$-projected component, which is determined by the Fe-O hybridization.
The $e_{2g}$ states are hybridized only with in-plane O 2$p_{x,y}$ orbitals while the $a_{1g}$ states are hybridized mostly with apical oxygen 2$p_z$ orbitals.
The $e_{1g}$ states are hybridized with both in-plane O 2$p_{x,y,z}$ and apical O 2$p_{x,y}$.
These 2$p$ characters in the hybridized Fe 3$d$ orbitals are picked up by the photon $\vec{E}$ vector parallel to the orbital directions.
Thus, the polarization dependence of the XAS spectra enables us to assign the orbital character of the Fe$^{2+}$ 3$d$ and Fe$^{3+}$ 3$d$ structures, as indicated in Fig.~5.
The assigned orbital energy structure is identical to that of LuFe$_2$O$_4$ in the previous study \cite{Ko}.

By combining all our observations, we can deduce one phenomenological result that an increase in the bilayer thickness, together with an elongation of the FeO$_5$ bipyramids in LuFe$_2$O$_4$ decreases the ferrimagnetic transition temperature T$_C$ without changing the Fe spin-ordering pattern.
Because the distance between the inter-layer Fe$^{2+}$ ions gets longer with increasing bilayer thickness, the direct exchange coupling between the inter-layer Fe$^{2+}$ ions also weakens, so the decrease in T$_C$ seem to be natural.
However, this cannot explain the invariance of the ferrimagnetic transition temperature in RFe$_2$O$_4$ (R=rare-earth elements) compounds \cite{Kimizuka}.
The experimental data show that the bilayer thickness does not affect the transition temperature in RFe$_2$O$_4$, which definitely does not agree with the result in this Sc-substituted LuFe$_2$O$_4$ case.
A more probable scenario is a decreased spin-orbit coupling effect in Fe$^{2+}$ ions.
Because of the elongation of the structure, the FeO$_5$ bipyramid becomes less asymmetric, as described above.
If the crystal electric field in FeO$_5$ has perfect $D_{3h}$ symmetry, the $e_{1g}$ doublet ($d_{yz/zx}=\frac{i}{\sqrt{2}}(|m_l=1\rangle+|-1\rangle)/\frac{1}{\sqrt{2}}(|1\rangle-|-1\rangle) $) is truly degenerate.
Then, the sixth 3$d$ electron in the Fe$^{2+}$ ion occupies the $d_{yz/zx}$ states equally, which has no orbital magnetic moment due to the perfect quenching if the spin-orbit coupling in Fe 3$d$ states is negligible.
In a real LuFe$_2$O$_4$ system, the spin-orbit coupling plays a key role in the ferromagnetic interaction between the Fe$^{2+}$ ions.
However, we note here that the asymmetry of FeO$_5$ with respect to the mirror $xy$-plane can contribute to the spin-orbit coupling effect.
Due to the asymmetry, the $e_{1g}$ state is split with unequal weights into $|m_l=\pm1\rangle$ components.
Usually, this effect is regarded as a negligible second-order perturbation, but it can be comparable to the spin-orbit coupling of the transition metal 3$d$ electrons when the asymmetry is quite large.
If LuFe$_2$O$_4$ is such a case, the polarized $|m_l\rangle$ components in the ground state increase the $\langle\vec{L}\cdot\vec{S}\rangle = \langle{L_z}\cdot{S_z}\rangle$ ($xy$-components are isotropic in this case) as well as the orbital angular momentum.
Consequently, the orbital moment quenching competes with the spin-orbit coupling effect.
Although Sc substitution does not recover $D_{3h}$ symmetry perfectly, it reduces the asymmetry in FeO$_5$.
Then, the increased orbital moment quenching reduces the spin-orbit coupling effect in the Fe$^{2+}$ ions, inducing a weakening of the ferromagnetic interaction between the Fe$^{2+}$ ions.

\section{Summary}

We successfully synthesized single crystals LuFe$_2$O$_4$ with Sc substituting for Lu ions by using a floating zone method and investigated the crystal structure distortion and the electronic/magnetic structure change by using XRD, magnetic susceptibility, XAS, and XMCD measurements.
The Rietveld structure analysis of the XRD patterns revealed that the Sc substitution induced an elongation of the FeO$_5$ bipyramidal cages in LuFe$_2$O$_4$ and an elongation of the unit cell.
At a concentration of Sc=0.3, a non-negligible decrease in the ferrimagnetic transition temperature is observed in the magnetic susceptibility curve, but the XAS/XMCD spectra does not change very much except for a small reduction of the dichroism signals at the Fe$^{3+}$ absorption edge.
The decrease in T$_C$ is quite noticeable because this class of materials is famous for having a robust transition temperature at 240 K.
We interpret the suppression of T$_C$ to be the result of the decreased spin-orbit coupling effect in the Fe$^{2+}$ $e_{1g}$ doublet under $D_{3h}$ symmetry, which is induced by the weakened structural asymmetry of the FeO$_5$ bipyramids in Sc-substituted LuFe$_2$O$_4$.

\begin{acknowledgments}
This work was supported by the National Research Foundation (NRF) of Korea grant funded by the Korean Government (Grant Nos. 2013R1A1A2058195 and 2010-0010771). The experiments at Pohang Light Source were supported in part by MSIP and POSTECH.
\end{acknowledgments}

\end{document}